\documentclass[sigconf, nonacm]{acmart}
\AtBeginDocument{%
  \providecommand\BibTeX{{%
    \normalfont B\kern-0.5em{\scshape i\kern-0.25em b}\kern-0.8em\TeX}}}

\usepackage{tabularx}
\usepackage{lscape}
\usepackage{stfloats}
\usepackage{multicol}
\usepackage{rotating}
\usepackage{multirow}
\usepackage{float}

\begin{document}

\title{Do Internal Software Metrics Have Relationship with Fault-proneness and Change-proneness?}

% \title{A Comparative Analysis of Code Smells between Open Source and Industrial systems}

\author{Md. Masudur Rahman}

% %\orcid{0000-0002-0931-1919}
\authornotemark[1]
\affiliation{ 
    \institution{Institute of Information Technology, University of Dhaka}
    \city{Dhaka}
    \country{Bangladesh}
    \postcode{1000}
}
\email{bit0413@iit.du.ac.bd}

\author{Toukir Ahammed}
\affiliation{
 	\institution{Institute of Information Technology, University of Dhaka}
	\city{Dhaka}
 	\country{Bangladesh}
 	\postcode{1000}
}
\email{toukir@iit.du.ac.bd}

 \author{Kazi Sakib}
 \affiliation{
 	\institution{Institute of Information Technology, University of Dhaka}
	\city{Dhaka}
 	\country{Bangladesh}
 	\postcode{1000}
}
\email{sakib@iit.du.ac.bd}

\renewcommand{\shortauthors}{Rahman et al.}
%\renewcommand\thesubsection{\Alph{subsection}}
%%
%% The abstract is a short summary of the work to be presented in the
%% article.

%\begin{CCSXML}
%<ccs2012>
%<concept>
%<concept_id>10010520.10010553.10010562</concept_id>
%<concept_desc>Computer systems organization˜Embedded systems</concept_desc>
%<concept_significance>500</concept_significance>
%</concept>
%<concept>
%<concept_id>10010520.10010575.10010755</concept_id>
%<concept_desc>Software and its engineering</concept_desc>
%<concept_significance>300</concept_significance>
%</concept>
%<concept>
%<concept_id>10010520.10010553.10010554</concept_id>
%<concept_desc>Software systems</concept_desc>
%<concept_significance>100</concept_significance>
%</concept>
%<concept>
%<concept_id>10003033.10003083.10003095</concept_id>
%<concept_desc>Software metrics</concept_desc>
%<concept_significance>100</concept_significance>
%</concept>
%</ccs2012>
%\end{CCSXML}
%\ccsdesc[500]{Software and its engineering}
%\ccsdesc{Software systems}
%\ccsdesc[100]{Software metrics}

\begin{abstract}
Fault-proneness is a measure that indicates the possibility of programming errors occurring within a software system. On the other hand, change-proneness refers to the potential for modifications to be made to the software. Both of these measures are crucial indicators of software maintainability, as they influence internal software metrics such as size, inheritance, and coupling, particularly when numerous changes are made to the system.
In the literature, research has predicted change- and fault-proneness using internal software metrics that is almost a decade old. However, given the continuous evolution of software systems, it is essential to revisit and update our understanding of these relationships. Therefore, we have conducted an empirical study to revisit the relationship between internal software metrics and change-proneness, and fault-proneness, aiming to provide current and relevant insights.
In our study, we identified 25 internal software metrics along with the measures of change-proneness and fault-proneness within the well-known open-source systems from the Apache and Eclipse ecosystems. We then analyzed the relationships between these metrics using statistical correlation methods. Our results revealed that most of the metrics have little to no correlation with fault-proneness. However, metrics related to inheritance, coupling, and comments showed a moderate to high correlation with change-proneness.
These findings will assist developers to minimize the higher correlated software metrics to enhance maintainability in terms of change- and fault-proneness. Additionally, these insights can guide researchers in developing new approaches for predicting changes and faults by incorporating the metrics that have been shown to have stronger correlations.
\end{abstract}


%%
%% This command processes the author and affiliation and title
%% information and builds the first part of the formatted document.
\maketitle

\section{Introduction}
\label{intro}
%%%%%%%%%%%%% Broad domain %%%%%%%%%%%%%%
Software quality is an important factor to measure how much a software system is maintainable. In ISO/IEC 9126 (International Organization for Standardization \cite{iso2001iec}) standard, software quality is described as a set of attributes (or metrics) of a software system by which its quality is described and evaluated. These quality attributes can be categorized into two groups \cite{alshayeb2009empirical}: (i) internal quality attributes -- attributes that can be directly measured from the source code of the software system (e.g., lines of code) and (ii) external quality attributes  -- attributes that can be measured indirectly from the software system (e.g., maintainability). By evaluating both internal and external quality attributes, developers can gain a comprehensive understanding of a software system's overall quality and its capacity for maintainability.

%%%%%%% Problem domain and motivation %%%%%%%
Internal quality attributes, referred to as internal software metrics in this study, serve as indicators of various aspects of software, including complexity, coupling, cohesion, abstraction, encapsulation, and documentation \cite{zhang2013does}. These metrics provide insight into different aspects of software maintainability. In contrast, external quality attributes such as change-proneness and fault-proneness are significant indicators used to assess a system's maintainability \cite{palomba2018diffuseness}. Improving maintainability is a primary objective for software developers during the maintenance phase of a system's life-cycle. To achieve this goal, it is essential to understand which internal software metrics significantly influence these maintainability attributes. By identifying and focusing on the internal metrics that impact change-proneness and fault-proneness, developers can strategically optimize their efforts to improve overall system maintainability. Therefore, this study seeks to determine the relationship between internal software metrics and the external attributes of change-proneness and fault-proneness. The findings will provide developers with valuable insights into which internal metrics are most impactful. By optimizing these key metrics, developers can enhance the maintainability of the system, thereby reducing its change-proneness and fault-proneness and making it more robust and easier to maintain.

%%%%%%%%% Existing work %%%%%%%%%%%
Many researches have been conducted by focusing on the empirical validation of the associations between internal software metrics and important maintainability metrics such as change-proneness and fault-proneness \cite{zhou2009examining}. Several works also investigated how change-proneness and fault-proneness can be predicted using the internal software metrics \cite{shatnawi2008effectiveness, lu2012ability, radjenovic2013software}. Rahman et al. \cite{rahman2013and} have shown that internal software metrics are not a good choice in defect prediction, rather process metrics such as commits, developers' characteristics, etc. are better at predicting defects. On the other hand, Lu et al. \cite{lu2012ability} showed that size-related metrics have a greater impact on change-proneness than other software metrics. However, most of this research has been performed since almost one decade ago. There has been a gap of many years since these relationships have been studied. In this paper, we aim to revisit the the relationship between internal software metrics and maintainability through change-proneness and fault-proneness in order to provide insights to developer communities regarding their effectiveness as indicators of maintainability. Our objective is to determine whether internal software metrics truly serve as reliable indicators or if they should be disregarded altogether.

%%%%%%%%%%%%%%%%% Methodology and Results %%%%%%%%%%%%%%%%
To identify the relationship between internal software metrics and maintainability attributes such as change-proneness and fault-proneness, an empirical study was conducted on 22 well-known open-source Java systems. From these systems, 25 internal software metrics were identified, encompassing various object-oriented code design aspects, including size, coupling, and inheritance. Then, a statistical correlation analysis was then performed to examine the relationships between each internal metric and both fault-proneness and change-proneness. The analysis revealed that many software metrics are significantly related to change-proneness, while most metrics show moderate to low correlation with fault-proneness. These findings provide valuable insights for software practitioners, enabling them to focus their efforts on the internal metrics that most impact change-proneness and fault-proneness, thereby improving the maintainability of their systems. Additionally, this study helps researchers to enhance the prediction models of change-proneness and fault-proneness, guiding future research and development in software engineering.

\section{Related Work}
There exist several works regarding the effect of internal software metrics with change-proneness and fault-proneness of a system \cite{malhotra2013investigation, olague2007empirical}. In this section, the existing literature is discussed.

Lu et al. \cite{lu2012ability} examined the ability of object-oriented (OO) metrics to predict software change-proneness. They showed that size-related metrics have a moderate effect in predicting change-proneness, while coupling-related metrics have a lower effect, and metrics related to inheritance have a poor effect. 

Zhou et al. \cite{zhou2009examining} showed that size-related class metrics have a confounding effect on the association between internal OO metrics such as cohesion, coupling, inheritance-related metrics, etc., and change-proneness. After conducting an empirical study on three open source systems, Eski et al. \cite{eski2011empirical} showed that there exist relations between metrics and change-proneness. However, significant metrics are different based on software domains. This leads to the necessity to re-investigate the relationship. 

According to Rahman et al. \cite{rahman2013and}, software metrics are important factors to predict fault-proneness. In this respect, careful choice of metrics is significant to improve the prediction accuracy. However, they showed that internal code-related metrics (e.g., size, complexity) are less useful in fault prediction whereas process-related metrics (e.g., number of changes, number of developers) are significant. Our investigation is to revisit their findings about whether the internal software metrics affect fault-proneness by analysing current software systems.

Several works have been performed where these two maintainability metrics -- fault-proneness and change-proneness are used to understand the relationship with code smells \cite{khomh2009exploratory, palomba2018diffuseness, rahman2023does}. They showed that smelly classes are more prone to changes and faults than non-smelly classes. 

In a systematic literature review, Radjenovic et al. \cite{radjenovic2013software} showed that object-oriented metrics are reported to be better at predicting fault-proneness than complexity and size metrics. In addition, coupling and size-related metrics are effective in fault prediction models \cite{aggarwal2009empirical}.

There exist a very little work on the relationship of software metrics with change-proneness and fault-proneness. Moreover, these works have been performed almost a decade ago. Since different change and fault prediction models use different software metrics, it is necessary to identify the relationship of internal software metrics with change-proneness and fault-proneness in the current software context. 

%%%%%%%%%% \section{Research Methodology}  %%%%%%%%%%
\section{Empirical Study Design}
The study aims to evaluate the relationship between each of the 25 internal software metrics, and change-proneness and fault-proneness. This evaluation will be conducted by analyzing how frequently these metrics appear in 22 well-known open-source software systems. The objective is to identify which specific metrics have a significant impact on the occurrence of changes and faults within the software.

\subsection{Formulating the Research Goal}
To investigate the relationship, the study particularly aims to answer the following two research questions:

\textbf{RQ1: \textit{What is the relationship between internal software metrics and change-proneness?}}

This research question aims to determine which internal software metrics are associated with change-proneness by examining their occurrences within various software systems. To identify these relationships, a correlation-based analysis is conducted, revealing whether these metrics are indeed related to change-proneness. The metrics that show a higher correlation with change-proneness are regarded as highly `impactful metrics' for this specific research question (RQ1). The findings from RQ1 will assist practitioners in focusing on these impactful metrics, ultimately aiding in the detection and reduction of change-proneness in software development and maintenance processes.

\textbf{RQ2: \textit{What is the relationship between internal software metrics and fault-proneness?}}

This research question examines which internal software metrics are related to fault-proneness by analyzing their occurrences within various software systems. Similar to the approach in RQ1, the goal is to identify metrics that show a strong correlation with fault-proneness, designating them as `impactful metrics' for this specific research question (RQ2). The findings from RQ2 will help practitioners focus on these significant metrics to detect and reduce fault-proneness in software development and maintenance.

\subsection{Systems under Study}
In order to conduct the empirical study and answer the research questions, we have analyzed 22 well-known open-source Java systems which belong to two major ecosystems: Apache\footnote{\url{https://www.apache.org}} and Eclipse\footnote{\url{https://www.eclipse.org}}.
Table \ref{tab:system} summarizes the analyzed systems, the latest stable versions, and their size in terms of the number of classes (NOCs), number of methods (NOMs), and lines of codes (LOCs). 
\begin{table}[!h]
	\centering
	\caption{Software systems involved in the study}
	\label{tab:system}
	\resizebox{0.47\textwidth}{!}{%
		\begin{tabular}{|l|l|l|l|l|l|} 
			\hline
			\begin{tabular}[c]{@{}l@{}}\textbf{Id }\\\textbf{No.}\end{tabular} & \textbf{Project} & \textbf{Version}   & \textbf{\#Classes} & \textbf{\#Methods} & \textbf{LOCs}        \\ 
			\hline
			1                                                                  & Activemq         & 5.17.0             & 4,747              & 42,349             & 421,979              \\ 
			\hline
			2                                                                  & Ant              & 1.9.0              & 1,595              & 13,768             & 138,391              \\ 
			\hline
			3                                                                  & Ant-ivy          & 2.5.0              & 814                & 7,540              & 76,678               \\ 
			\hline
			4                                                                  & Cassandra        & 4.0.0              & 8,101              & 76,237             & 1,483,583            \\ 
			\hline
			5                                                                  & Cayenne          & 4.1                & 6,656              & 41,339             & 485,781              \\ 
			\hline
			6                                                                  & CXF              & 3.5.0              & 11,109             & 103,490            & 1,120,148            \\ 
			\hline
			7                                                                  & Drill            & 1.10.0             & 6,414              & 60,203             & 533,478              \\ 
			\hline
			8                                                                  & Hadoop           & 3.3.0              & 17,393             & 143,785            & 1,762,363            \\ 
			\hline
			9                                                                  & HBase            & 3.0.0              & 9,457              & 71,619             & 823,856              \\ 
			\hline
			10                                                                 & Hive             & 3.1.0              & 14,466             & 107,711            & 1,257,002            \\ 
			\hline
			11                                                                 & Incubator-livy   & 0.7.0              & 201                & 811                & 8,069                \\ 
			\hline
			12                                                                 & Jackrabbit       & 2.9.0              & 3,209              & 29,053             & 332,826              \\ 
			\hline
			13                                                                 & Jena             & 4.5.0              & 7,077              & 66,138             & 576,575              \\ 
			\hline
			14                                                                 & Karaf            & 4.4.0              & 1,760              & 10,332             & 126,431              \\ 
			\hline
			15                                                                 & Lucene           & 9.3.0              & 7,501              & 52,968             & 827,867              \\ 
			\hline
			16                                                                 & Nutch            & 2.4                & 519                & 3,136              & 38,268               \\ 
			\hline
			17                                                                 & Pig              & 0.10.0             & 2,222              & 15,433             & 231,371              \\ 
			\hline
			18                                                                 & Poi              & 5.2.2              & 4,291              & 40,032             & 413,657              \\ 
			\hline
			19                                                                 & Qpid             & 0.3                & 3,374              & 29,189             & 327,821              \\ 
			\hline
			20                                                                 & Struts           & 6.0.0              & 2,378              & 18,640             & 186,595              \\ 
			\hline
			21                                                                 & Wicket           & 7.2.0              & 4,402              & 20,678             & 215,286              \\ 
			\hline
			22                                                                 & Xerces           & 2.9.0              & 800                & 9,809              & 131,326              \\ 
			\hline
			& \multicolumn{2}{l|}{\textbf{Total}}   & \textbf{118,486}   & \textbf{964,260}   & \textbf{11,519,351}  \\ 
			\hline
			& \multicolumn{2}{l|}{\textbf{Average}} & \textbf{5,385}     & \textbf{43,830}    & \textbf{523,606}    \\
			\hline
		\end{tabular}
	}
\end{table}

These systems have been selected based on several criteria -- (i) The systems from these ecosystems are highly regarded in the software engineering research domain, as evidenced by studies such as those by Palomba et al. \cite{palomba2018diffuseness} and Pecorelli et al. \cite{pecorelli2020developer}. (ii) These systems utilize Bugzilla \footnote{\url{http://www.bugzilla.org}.} or Jira\footnote{\url{https://www.atlassian.com/software/jira}.} as issue trackers, which are crucial for identifying fault-related information. (iii) The selected systems vary significantly in size, ranging from 8,069 to 1,762,363 LOCs. On average, they have 5,385 classes, 43,830 methods, and 523,606 lines of code, making them large enough to provide robust data for our analysis.

\subsection{Selection of Internal Software Metrics}
\label{InternalMetrics}
For this study, we consider 25 internal quality attributes and refer these attributes as internal software metrics. The list of the metrics with a short description is shown in Table \ref{Tab_results}. We choose these metrics for the study because -- (i) these cover six factors of an object oriented software system, that is, complexity, coupling, cohesion, abstraction, encapsulation, and documentation, and (ii) these factors can measure different aspects of software maintainability \cite{zhang2013does}. These metrics are computed by a well-known tool designed for research purposes, called Understand\footnote{\url{https://www.scitools.com/}} \cite{zhang2013does}.

\begin{table*}[!h]
	\centering
	\caption{Relationship of Internal Software Metrics with Change-proneness and Fault-proneness}
	\label{Tab_results}
	\resizebox{1\textwidth}{!}{%
	\begin{tabular}{|l|l|l|c|c|} 
		\hline
		\multicolumn{1}{|c|}{\textbf{Id No.}} & \multicolumn{1}{c|}{\textbf{Software Metric}} & \multicolumn{1}{c|}{\textbf{Description}}                                                                                                  & \begin{tabular}[c]{@{}c@{}}\textbf{Correlation with }\\\textbf{ Change-proneness,}\\\textbf{ $r_{cp}$}\end{tabular} & \begin{tabular}[c]{@{}c@{}}\textbf{Correlation with }\\\textbf{ Fault-proneness,}\\\textbf{ $r_{fp}$}\end{tabular}  \\ 
		\hline
		1                                     & CountDeclMethod, NOM                          & Number of local methods                                                                                                                    & \textbf{0.50}                                                                                                       & \textit{0.35}                                                                                                       \\ 
		\hline
		2                                     & CountDeclMethodDefault, NDM                   & Number of local default methods                                                                                                            & \textit{0.47}                                                                                                       & 0.22                                                                                                                \\ 
		\hline
		3                                     & CountDeclMethodPrivate, NPriM                 & Number of local private methods                                                                                                            & \textit{0.46}                                                                                                       & 0.25                                                                                                                \\ 
		\hline
		4                                     & CountDeclMethodProtected, NProM               & Number of local protected methods                                                                                                          & \textit{0.32}                                                                                                       & 0.15                                                                                                                \\ 
		\hline
		5                                     & CountDeclMethodPublic, NPM                    & Number of local public methods                                                                                                             & \textbf{0.52}                                                                                                       & \textit{0.37}                                                                                                       \\ 
		\hline
		6                                     & CountStmt, NOS                                & Number of statements                                                                                                                       & \textit{0.46}                                                                                                       & \textit{0.33}                                                                                                       \\ 
		\hline
		7                                     & CountStmtDecl, NDS                            & Number of declarative statements                                                                                                           & \textit{0.45}                                                                                                       & \textit{0.31}                                                                                                       \\ 
		\hline
		8                                     & CountStmtExe, NexS                            & Number of executable statements                                                                                                            & \textit{0.46}                                                                                                       & \textit{0.32}                                                                                                       \\ 
		\hline
		9                                     & SumCyclomatic, CC                             & Sum of cyclomatic complexity of all nested methods                                                                                         & \textit{0.48}                                                                                                       & \textit{0.33}                                                                                                       \\ 
		\hline
		10                                    & LCOM                                          & Lack of cohesion in methods                                                                                                                & \textit{0.49}                                                                                                       & \textit{0.33}                                                                                                       \\ 
		\hline
		11                                    & Depth of Inheritance Tree, DIT                & Maximum depth of class in inheritance tree                                                                                                 & \textit{0.41}                                                                                                       & \textit{0.32}                                                                                                       \\ 
		\hline
		12                                    & IFANIN                                        & Number of immediate base classes                                                                                                           & \textbf{0.51}                                                                                                       & \textit{0.38}                                                                                                       \\ 
		\hline
		13                                    & Coupling Between Objects, CBO                 & \begin{tabular}[c]{@{}l@{}}Number of other classes to which a particular class \\ is coupled\end{tabular}                                  & \textbf{0.51}                                                                                                       & \textit{0.36}                                                                                                       \\ 
		\hline
		14                                    & Number of Children, NOCh                      & Number of immediate subclasses                                                                                                             & \textit{0.32}                                                                                                       & 0.26                                                                                                                \\ 
		\hline
		15                                    & Response For a Class, RFC                     & Number of methods including inherited ones                                                                                                & \textit{0.32}                                                                                                       & \textit{0.30}                                                                                                       \\ 
		\hline
		16                                    & Number of Instance Methods, NIM               & \begin{tabular}[c]{@{}l@{}}Number of methods defined in a class that are only \\ accessible through an object of that class\end{tabular}   & \textit{0.49}                                                                                                       & \textit{0.32}                                                                                                       \\ 
		\hline
		17                                    & Number of Instance Variables, NIV             & \begin{tabular}[c]{@{}l@{}}Number of variables defined in a class that are only \\ accessible through an object of that class\end{tabular} & \textit{0.48}                                                                                                       & 0.29                                                                                                                \\ 
		\hline
		18                                    & Weighted Methods per Class, WMC               & Sum of the complexities of all class methods                                                                                               & \textit{0.46}                                                                                                       & 0.29                                                                                                                \\ 
		\hline
		19                                    & Classes, NOC                                  & Number of Classes                                                                                                                          & \textit{0.49}                                                                                                       & \textit{0.36}                                                                                                       \\ 
		\hline
		20                                    & Files, NOF                                    & Number of Files                                                                                                                            & \textit{0.38}                                                                                                       & 0.23                                                                                                                \\ 
		\hline
		21                                    & Lines, NL                                     & Number of all lines                                                                                                                        & \textit{0.47}                                                                                                       & \textit{0.32}                                                                                                       \\ 
		\hline
		22                                    & Lines Blank, BLOC                             & Number of blank lines of code                                                                                                              & \textbf{0.50}                                                                                                       & \textit{0.33}                                                                                                       \\ 
		\hline
		23                                    & Lines of Code, LOC                            & Number of lines containing source code                                                                                                     & \textit{0.46}                                                                                                       & \textit{0.32}                                                                                                       \\ 
		\hline
		24                                    & Lines Comment, NCL                            & Number of comment lines of code                                                                                                            & 0.25                                                                                                                & 0.12                                                                                                                \\ 
		\hline
		25                                    & RatioCommentToCode, RCTC                      & Ratio of comment lines to code lines                                                                                                       & \textit{-0.47}                                                                                                      & \textit{-0.42}                                                                                                      \\
		\hline
	\end{tabular}	
	}
	\\{\raggedright \vspace{0.2cm} \small [* Bold, italic and normal fonts indicate high, moderate and low (or no) correlation respectively.]\par}
\end{table*}

\subsection{Detection of Change-proneness}
\label{DetectCP}
The change-proneness of a class is computed as the total number of changes between two releases \cite{palomba2018diffuseness}. Both addition and deletion are considered as changes to a class. So, the number of changes of a class $C_i$ is calculated using Equation \ref{eq:cp_formula}.
\begin{equation}
	\#changes(C_i) = added(C_i) + deleted(C_i)
	\label{eq:cp_formula}
\end{equation}
where $added(C_i)$ refers to the number of added lines and $deleted(C_i)$ refers to the number of deleted lines between two consecutive releases. To compute $added(C_i)$) and ($deleted(C_i)$) of each class, the commit history is analyzed from the version control system.

\subsection{Detection of Fault-proneness}
\label{DetectFP}
The fault-proneness of a class is measured as the number of bug-fixing changes in a class, as after fixing a fault, it can be detected from the source code. These changes include both addition and deletion statements in source code through commits which are referred as bug-fixing commits. Bug-fixing commits are identified by searching commits that contain \textit{Issue ID} or \textit{Bug ID} in commit messages. The search is performed using regular expression. For example,  the following commit message from \textit{Cayenne} project \footnote{https://github.com/apache/cayenne} can be identified with the regular expression ``\textit{CAY-\textbackslash \textbackslash d+}'': 
\begin{quote}
	``CAY-2732 Exception when creating ObjEntity from a DbEntity''
\end{quote}

Using \textit{Issue ID} found in commit messages, the corresponding issue report is extracted from the issue tracker such as \textit{Jira} and \textit{Bugzilla}. Issues related to bugs are separated, i.e., the type of issue is \textit{bug},  excluding \textit{enhancement} type issues. To exclude duplicated or false-positive bugs that can bias the result, only bugs that have the status \textit{Closed} or \textit{Resolved} and the resolution \textit{Fixed} are considered \cite{palomba2018diffuseness}. Finally, the number of bug-fixing changes of a class $C_i$ is calculated as the sum of added and deleted lines only through bug-fixing commits using Equation \ref{eq:fp_formula}.
\begingroup
\fontsize{8pt}{10pt}\selectfont
\begin{equation}
	\#bug\_fixing\_changes(C_i) = fix\_added(C_i) + fix\_deleted(C_i) 
	\label{eq:fp_formula}
\end{equation}
\endgroup
Here, $fix\_added(C_i)$ and $fix\_deleted(C_i)$ are the number of added and deleted lines respectively for fixing changes.

\subsection{Data Analysis}
\label{DataAnalysis}
To conduct the data analysis, we have used the Spearman's Rank Correlation Coefficient \cite{gupta2020fundamentals} for each of the software metrics with respect to change-proneness and fault-proneness of the analysed systems as we observe the trend is monotonic between these two variables. A monotonic trend suggests that both variables exhibit parallel increases or decreases in the same or the opposite direction, but not necessarily at a constant rate a linear relationship. For instance, we have taken the score of metric \textit{Lines of Code (LOC)} of each system as the \textit{x} variable and change-proneness of the corresponding system as the \textit{y} variable to measure the correlation coefficient between them. Similarly, we have taken \textit{LOC} and fault-proneness to measure the relationship between them. In this way, we have calculated the relationships for all the metrics with change- and fault-proneness.

Moreover, to interpret the correlation coefficient, we have followed the guidelines provided by Cohen \cite{cohen2013statistical}. It is considered that there is no correlation when \textit{0 $\leq$ r $<$ 0.1}, small (or low) correlation when \textit{0.1 $\leq$ r $<$ 0.3}, medium (or moderate) correlation when \textit{0.3 $\leq$ r $<$ 0.5}, and strong (or high) correlation when \textit{0.5 $\leq$ r $\leq$ 1}. 

The source code regarding change- and fault-proneness detection and dataset including metrics' score used in the study have been given here\footnote{\url{https://tinyurl.com/47trdebt}} for replication and further research purposes.

%%%%%%%%% End of Empirical Study Design %%%%%%%%%%%%%%%%

\section{Result Analysis}
\label{results}
In this section, we delve into the findings of our study, which examines the relationships between each of the internal software metrics and two key aspects: change-proneness and fault-proneness. A comprehensive presentation of these detailed results is provided in Table \ref{Tab_results}.
\\
\\
\textbf{Relationship between Internal Software Metrics and Change-proneness (RQ1).}

Table \ref{Tab_results} provides a comprehensive analysis of the correlation between various internal software metrics and change-proneness. The correlation coefficients ($r_{cp}$) listed in Table \ref{Tab_results} indicate the strength of the relationship between each metric and the tendency of the software to undergo changes.
Based on the correlation coefficients ($r_{cp}$) from Table \ref{Tab_results}, the software metrics can be categorized into three impactful groups: high, moderate, and low impact having high, moderate and low (or no) correlation with change-proneness respectively.

\begin{enumerate}
\item[1.] \textbf{High Impactful Metrics (0.5 $\leq$ $r_{cp}$ $\leq$ 1):} These metrics have a strong relationship with change-proneness, indicating that classes with higher values for these metrics are significantly more likely to undergo changes. 5 metrics fall into this group: \textit{CountDeclMethod (Number of local methods, NOM); CountDeclMethodPublic (Number of local public methods, NPM); IFANIN (Number of immediate base classes); Coupling Between Objects (CBO)}; and \textit{Lines Blank	(Number of blank lines of code, BLOC)}.

\item[2.] \textbf{Moderate Impactful Metrics (0.3 $\leq$ $r_{cp}$ $<$ 0.5):} These metrics show a moderate relationship with change-proneness, suggesting a noticeable but less strong influence on the likelihood of changes. 19 metrics fall into this group: \textit{CountDeclMethodDefault	(Number of local default methods, NDM); CountDeclMethodPrivate (Number of local private methods, NPriM); CountDeclMethodProtected	(Number of local protected methods, NProM); CountStmt (Number of statements, NOS); CountStmtDecl (Number of declarative statements, NDS); CountStmtExe (Number of executable statements, NexS); SumCyclomatic (Sum of cyclomatic complexity of all nested methods, NexS); LCOM (Lack of cohesion in methods); Depth of Inheritance Tree (Maximum depth of class in inheritance tree, DIT); Number of Children (Number of immediate subclasses, NOCh); Response For a Class	(Number of methods including inherited ones, RFC); Number of Instance Methods	(Number of methods defined in a class that are only accessable through an object of that class, NIM); Number of Instance Variables (Number of variables defined in a class that are only accessable through an object of that class, NIV); Weighted Methods per Class (Sum of the complexities of all class methods, WMC); Classes (Number of Classes, NOC); Files	(Number of Files, NOF); Lines (Number of all lines, NL); Lines of Code (Number of lines containing source code, LOC);} and \textit{RatioCommentToCode (Ratio of comment lines to code lines, RCTC)}.

\item[3.] \textbf{Low Impactful Metrics (0.0 $\leq$ $r_{cp}$ $<$ 0.3):} Only one metric falls into this group: \textit{Lines Comment	(Number of comment lines of code, NCL)}. This metric has a low relationship with change-proneness, indicating a weaker influence on the likelihood of changes.
\end{enumerate}

Only one metric \textit{RatioCommentToCode (RCTC), $r_{cp}=-0.47$} has a moderate but negative correlation. That is, if comments increase in a system, change-proneness decreases and vice-versa. It could be possible that comments help in understanding source code and thus resisting to code changes. So developers should focus on this metric to reduce change-proneness. 

Overall, the analysis shows that all of the metrics except \textit{Lines Comment} have moderate to strong correlations with change-proneness. Metrics related to methods, inheritance and coupling tend to have higher correlations, indicating their significant impact on the likelihood of a class undergoing changes. These findings are essential for developers and software practitioners as these highlight which aspects of the codebase are more prone to changes and, therefore, need more attention during the development and maintenance phases. However, these findings contrast the previous findings conducted by Lu et al. \cite{lu2012ability}, where they showed size, coupling and inheritance-related metrics have moderate, lower and poor predictive ability of change-proneness respectively.
\\
\\
\textbf{Relationship between Internal Software Metrics and Fault-proneness (RQ2).}

Table \ref{Tab_results} provides a comprehensive analysis of the correlation between various internal software metrics and fault-proneness. The correlation coefficients ($r_{fp}$) listed in Table \ref{Tab_results} indicate the strength of the relationship between each metric and the tendency of the software to undergo changes.
Based on the correlation coefficients ($r_{fp}$) from Table \ref{Tab_results}, the software metrics can be categorized into three impactful groups: high, moderate, and low impact having high, moderate and low (or no) correlation with change-proneness respectively.

\begin{enumerate}
	\item[1.] \textbf{High Impactful Metrics (0.5 $\leq$ $r_{fp}$ $\leq$ 1):} These metrics have a strong relationship with faults-proneness, indicating that classes with higher values for these metrics are significantly more likely to undergo occurring faults. Interestingly, no metrics have been found in this group.
	
	\item[2.] \textbf{Moderate Impactful Metrics (0.3 $\leq$ $r_{fp}$ $<$ 0.5):} These metrics show a moderate relationship with fault-proneness, suggesting a noticeable but less strong influence on the likelihood of faults. 17 metrics fall into this group: \textit{CountDeclMethod (Number of local methods, NOM); CountDeclMethodPublic (Number of local public methods, NPM); CountStmt (Number of statements, NOS); CountStmtDecl (Number of declarative statements, NDS); CountStmtExe (Number of executable statements, NexS); SumCyclomatic (Sum of cyclomatic complexity of all nested methods, NexS); LCOM (Lack of cohesion in methods); Depth of Inheritance Tree (Maximum depth of class in inheritance tree, DIT); IFANIN (Number of immediate base classes); Coupling Between Objects (CBO); Response For a Class	(Number of methods including inherited ones, RFC); Number of Instance Methods	(Number of methods defined in a class that are only accessible through an object of that class, NIM); Classes (Number of Classes, NOC); Lines (Number of all lines, NL); Lines Blank	(Number of blank lines of code, BLOC); Lines of Code (Number of lines containing source code, LOC);} and \textit{RatioCommentToCode (Ratio of comment lines to code lines, RCTC)}.	
	
	\item[3.] \textbf{Low Impactful Metrics (0.0 $\leq$ $r_{fp}$ $<$ 0.3):} These metrics have a low relationship with fault-proneness, indicating a weaker influence on the likelihood of faults. 8 metrics fall into this group: \textit{CountDeclMethodDefault	(Number of local default methods, NDM); CountDeclMethodPrivate (Number of local private methods, NPriM); CountDeclMethodProtected	(Number of local protected methods, NProM); Number of Children (Number of immediate subclasses, NOCh); Number of Instance Variables (Number of variables defined in a class that are only accessible through an object of that class, NIV); Weighted Methods per Class (Sum of the complexities of all class methods, WMC); Files	(Number of Files, NOF); Lines Comment	(Number of comment lines of code, NCL)}.
\end{enumerate}

\textit{RatioCommentToCode (RCTC)} metric have moderate but negative correlations with fault-proneness as like change-proneness. That is, comments reduce fault-proneness and therefore developers should focus on this particular software metric to develop maintainable software systems. 

Overall, the analysis shows that all of the metrics have low to moderate correlations with fault-proneness. Metrics related to methods, statements, inheritance, cohesion and coupling tend to have higher correlations, indicating their significant impact on the likelihood of a class occurring faults. On the other hand, data or variable and some method related metrics have low correlations. These findings are essential for developers and software practitioners as these highlight which aspects of the codebase are more prone to faults and, therefore, need more attention during the development and maintenance phases. However, our finding supports the previous finding observed by Rahman et al. \cite{rahman2013and} where they showed that internal software metrics have lower fault predictive ability. 

%\noindent\fbox{
%\parbox{0.5\textwidth}{
%	\textbf{Summary for RQ1 and RQ2.} We have observed that internal software metrics have no or low impact on fault-proneness except \textit{comments}. On the other hand, inheritance, coupling and comments related metrics have moderate or high impact on change-proneness, whereas most of the size related metrics have low impact.
%}}

\section{Implications of the Result}
The study's findings on the correlation between software metrics and both change-proneness and fault-proneness have significant implications for software development and maintenance practices. Here are the key implications:

\begin{enumerate}
	\item[1.] Metrics with high correlation with change-proneness highlight critical areas in the source-code that are more likely to require changes. Addressing these areas can help stabilize the code and reduce the frequency of changes. Therefore, it is necessary to focus on classes with a high number of public methods (NPM) and high coupling (CBO) as they are more prone to changes. Also, it is important to implement design reviews and refactoring to reduce coupling and stabilize public interfaces.
	
	\item[2.] Understanding which metrics are strongly correlated with change-proneness and fault-proneness allows development teams to allocate resources more effectively. High-priority areas can receive more attention in terms of testing, review, and refactoring.
	
	\item[3.] Focusing on metrics that correlate with higher change-proneness and fault-proneness can lead to improvements in overall code quality, making the software more maintainable and less prone to defects.
	
	\item[4.] The significant correlation between documentation-related metrics such as ratio of comments to code (RCTC) and both change- and fault-proneness underscores the need for thorough and consistent documentation, which can help reduce the likelihood of introducing faults during maintenance and development. This can be achieved by setting documentation standards and incorporating documentation as a part of the development process. Use code reviews to ensure that comments are comprehensive and useful, which can help in understanding the code better and reducing change- and fault-proneness.
	
	\item[5.] The findings help developers to understand the importance of the internal software metrics and their impact on software quality resulting in writing maintainable code, monitoring and reviewing code, and maintaining documentation.
	
	\item[6.] The study finds that while internal software metrics generally have a low impact on fault-proneness, metrics related to comments, inheritance, and coupling are more impactful on change-proneness. These findings provide valuable insights for software practitioners aiming to optimize maintainability and for researchers developing predictive models for change- and fault-proneness.	
	
\end{enumerate}

By leveraging the correlations between software metrics, change-proneness, and fault-proneness, developers can make more informed decisions that enhance software quality and stability. Prioritizing high-risk areas, enhancing documentation, adopting a holistic approach to metrics, implementing robust testing strategies, continuously monitoring, and educating developers are key steps towards achieving these goals. These practices not only help in managing and mitigating risks but also contribute to creating a more maintainable and reliable codebase.
	
\section{Threats to Validity}
\label{Threats}
The potential threats to external validity have been identified in this study while generalizing the findings of the study. First, we have used Java systems in our study, and there is a possibility that the results would be different for other object-oriented languages, like - C\#. Second, we have used 25 internal software metrics, and the results cannot be generalized to other software metrics. Finally, we cannot extrapolate our results to other open-source and industrial systems.

The potential threats to internal validity concern factors that could influence our observations. We are aware that we cannot claim a direct cause-effect relationship between the internal software metrics, and software change- and fault-proneness. In particular, our observations may be influenced by the different factors related to development phases (e.g., experience of developers, workload, etc.). However, we have focused only on the source code-related metrics of software and other external factors are out of the scope of this study.

\section{Conclusion}
\label{conclusion}
This study has provided critical insights into the relationships between internal software metrics and their impact on change-proneness and fault-proneness within Java systems. The findings highlight that certain metrics, such as \textit{CountDeclMethod (NOM), CountDeclMethodPublic (NPM), IFANIN,} and \textit{Coupling Between Objects (CBO)}, show a high correlation with change-proneness, while documentation-related metrics, particularly the \textit{RatioCommentToCode (RCTC)}, exhibit a moderate negative correlation with both change- and fault-proneness. Our findings help software practitioners to optimize the impactful metrics to enhance software maintainability in terms of change- and fault-proneness. In addition, these findings help researchers to use impactful internal software metrics to detect these two maintainability metrics. As for future research, we will focus on other maintainability aspects such as understandability, testability, etc. to measure their impact on the change- and fault-proneness.

%\vspace{-3mm}
%\begin{acks}
%We gratefully acknowledge the support of Information and Communication Technology (ICT) Division, Ministry of Posts, Telecommunications and Information Technology, Bangladesh through Grant/Fellowship Number: 56.00.0000.052.33.005.21-2, 18-01-2022.
%\end{acks}

%%
%% The next two lines define the bibliography style to be used, and
%% the bibliography file.
\balance

\bibliographystyle{ACM-Reference-Format}
\bibliography{sample-base}

\end{document}